\documentclass[%
 reprint, %'mimics final production'
showpacs,
 amsmath,amssymb,
 aps,
prl,
floatfix,
 lengthcheck,
]{revtex4-1}

\usepackage{amsmath, amsthm, amssymb}
\usepackage{graphicx}

\newcommand{\HOLU}{{\sc HOMO-LUMO} }
\newcommand{\LU}{{\sc LUMO} }
\newcommand{\HO}{{\sc HOMO} }

\renewcommand\Re{\operatorname{Re}}
\renewcommand\Im{\operatorname{Im}}
\newcommand{\Tr}[1]{\operatorname{Tr}\left\{ #1 \right\}}

\begin{document}

\title{The number of transmission channels through a single-molecule junction}

\author{J.~P.~Bergfield}
\affiliation{College of Optical Sciences, University of Arizona, 1630 East University Boulevard, AZ 85721}
\email{justinb@email.arizona.edu}
\author{J.~D.~Barr} 
\affiliation{Department of Physics, University of Arizona, 1118 East Fourth Street, Tucson, AZ 85721}

\author{C.~A.~Stafford} 
\affiliation{Department of Physics, University of Arizona, 1118 East Fourth Street, Tucson, AZ 85721}

\begin{abstract}
We calculate transmission eigenvalue distributions for Pt--benzene--Pt and Pt--butadiene--Pt junctions
using realistic state-of-the-art many-body techniques.  An effective field theory of interacting $\pi$-electrons %using many-body theory.
is used to include screening and van der Waals interactions with the metal electrodes.
%A semi-empirical model of the electronic structure is utilized that accurately describes $\pi$-conjugation and Coulomb interactions,
%includes screening from the metal electrodes.
We find that the number of dominant transmission channels
in a molecular junction is equal to the degeneracy of the molecular orbital closest to the metal Fermi level.
\end{abstract}

\maketitle

%{\bf keywords:}
%many-body theory, transmission channels, single-molecule junction, lead-molecule coupling, wave-particle duality
%\end{keywords}

% Introduction

The transmission eigenvalues $\tau_n$  constitute a {\em mesoscopic PIN code} \cite{Cron01} %fully 
characterizing quantum transport through any nanoscale device.
For a single-atom contact between two metallic electrodes, the number of transmission channels 
is simply given by the chemical valence of the atom \cite{Scheer98}.
Recently, highly-conductive single-molecule junctions (SMJ) with multiple
transport channels have been formed from benzene molecules between Pt electrodes \cite{Kiguchi08}.
This raises the question if there exists a similarly {\em simple criterion determining the number of
transmission eigenchannels in a SMJ.}  

Previous calculations \cite{Heurich02,Solomon06b} using effective single-particle models based on density functional theory
appear to answer the above question in the negative.  %indicate that such an analysis is quite complicated 
However, we find that transmission channel distributions calculated using many-body theory do yield a simple, intuitive answer to this
important question.

%The transmission eigenvalues of a SMJ 
For a two-terminal SMJ,
$\tau_n$ are eigenvalues of the elastic transmission matrix \cite{footnote_Tmatrix,Heurich02,Solomon06b,Paulsson07}
\begin{equation}
%T_{\alpha\beta}(E)=G(E) \Gamma^\beta(E) G^\dagger(E) \Gamma^\alpha(E),
T(E)=\Gamma_1(E) G(E) \Gamma_2(E) G^\dagger(E),
\label{eq:trans_matrix}
\end{equation}
where $G$ is the retarded Green's function \cite{Bergfield09} of the SMJ and $\Gamma_\alpha$ is the tunneling-width matrix describing
the coupling of the molecule to lead $\alpha$.  The number of transmission
channels is equal to the rank of the matrix (1), which is in turn limited by the ranks of the matrices $\Gamma_\alpha$ and $G$.  The additional two-fold spin degeneracy of
each resonance is considered implicit.  
The rank of $\Gamma_\alpha$ is equal to the number of covalent bonds formed between the molecule 
and lead $\alpha$.  Thus, for example, in a SMJ where a Au electrode bonds to an organic molecule via a thiol group, only a single bond is formed, and there
is only one non-negligible transmission channel \cite{Djukic06,Solomon06b}.  
In Pt--benzene--Pt junctions, however, each Pt electrode forms multiple bonds
to the benzene molecule \cite{Kiguchi08}.

\begin{figure}[tb]
	\centering
		\includegraphics[width=3.3in]{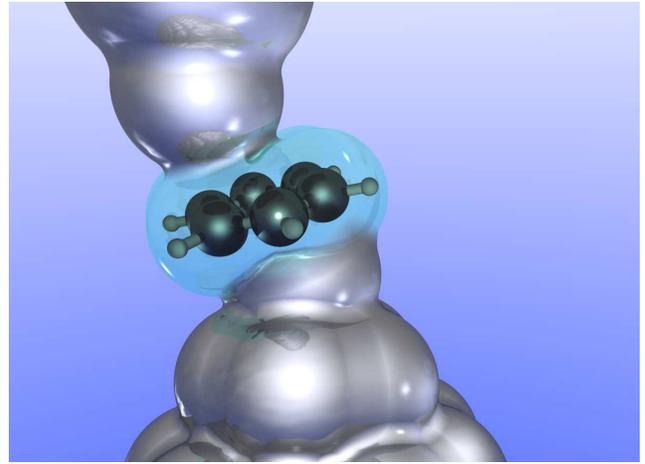}
\caption{Schematic diagram of a Pt--benzene--Pt junction.  The lead-molecule coupling is mediated predominantly 
by individual Pt atoms at the tips of each electrode.
}
	\label{fig:multimode_paper_benzene_Pt}
\end{figure}

In this article, we investigate how transmission eigenvalue distributions of SMJs depend on the number of lead-molecule bonds and on 
molecular symmetry using a many-body theory of transport \cite{Bergfield09}.  Specifically, we focus on junctions
 with benzene (${\rm C_6H_6}$) and butadiene (${\rm C_4H_6}$) bonded to two Pt leads (see Fig.\ \ref{fig:multimode_paper_benzene_Pt}). 
Consistent with %the findings of 
refs.\ \citenum{Kiguchi08} and \citenum{Solomon06b}, we find that the total number of {\em nonzero} transmission eigenvalues in a SMJ is limited
only by the number of bonds to each electrode.  However, increasing the number of bonds past a certain point leads to additional channels with very small
transmission $\langle \tau_n\rangle \ll 1$.  The central finding of this article is that in SMJs with sufficient numbers of lead-molecule bonds 
{\em the number of dominant transmission channels
%, defined as channels with transmission much greater than all others, 
is equal to the degeneracy of the molecular orbital closest to the metal Fermi level}.  Additional transmission channels stemming from further off-resonant
molecular states are strongly suppressed, but may still be experimentally resolvable \cite{Kiguchi08} for very strong lead-molecule hybridization.

%%%%%%%%%%%%%%%%%%%%%%%%%%%%%%%%%%%%%%%%%%%%%%%%
% Background/description of our many-body theory
%%%%%%%%%%%%%%%%%%%%%%%%%%%%%%%%%%%%%%%%%%%%%%%%%
\section{Many-body Theory of Transport}
When macroscopic leads are attached to a single molecule, a
SMJ is formed, transforming the few-body molecular problem into a true 
many-body problem. 
Until recently, a theory of transport in SMJs that properly accounts for both the
{\em particle and wave character} of the electron has been lacking, so that the
Coulomb blockade and coherent transport regimes were considered 
`complementary.'\cite{Geskin09} 
In this article, we utilize a nonequilibrium many-body theory \cite{Bergfield09} that 
correctly accounts for wave-particle duality, 
reproducing the key features of both the Coulomb blockade and coherent transport regimes. 
Previous applications \cite{Bergfield09,Bergfield09b,Bergfield2010,Bergfield2010b} %of our many-body theory 
to SMJs utilized a semi-empirical Hamiltonian \cite{Castleton02}
for the $\pi$-electrons, which accurately describes the gas-phase spectra of conjugated organic molecules.  This approach 
should be adequate to describe molecules weakly coupled to metal electrodes, e.g.\ via thiol linkages.  However, in junctions
where the $\pi$-electrons bind directly to the metal electrodes \cite{Kiguchi08}, 
the lead-molecule coupling may be so strong that the molecule itself is significantly
altered, necessitating a more fundamental molecular model.
To address this issue, we have developed an {\em effective field theory of interacting $\pi$-electrons}%\cite{JoshUnpublished}
, in which the form of the molecular Hamiltonian is derived from symmetry principles and electromagnetic theory (multipole expansion),
rather than using an ad-hoc parametrization \cite{Castleton02}.
The resulting formalism constitutes a state-of-the-art many-body theory that provides a realistic description of lead-molecule
hybridization and van der Waals coupling, as well as the screening of intramolecular interactions by the metal electrodes, all of
which are essential for a quantitative description of strongly-coupled SMJs \cite{Kiguchi08}.

%%%%%%%%%%%%%%%%%%
% Transport basics
%%%%%%%%%%%%%%%%%%

The Green's function of a SMJ has the form \cite{Bergfield09}
\begin{equation}
G(E) = \left[ G_{\rm mol}^{-1}(E) - \Sigma_{\rm T}(E) - \Delta \Sigma_{\rm C}(E)\right]^{-1},
\label{eq:Dyson2}\end{equation}
where $G_{\rm mol}$ is the molecular Green's function,
% in the limit of infinitessimal tunneling widths (sequential-tunneling limit),
$\Sigma_{\rm T}$ is the tunneling self-energy matrix, whose imaginary part is given by $\Im \Sigma_{\rm T} = -\sum_\alpha \Gamma_\alpha/2$, and
$\Delta \Sigma_{\rm C}$ is the correction to the Coulomb self-energy due to the broadening of the molecular resonances in the junction.
At room temperature and for small bias voltages, $\Delta \Sigma_{\rm C}\approx 0$ in the cotunneling regime \cite{Bergfield09} 
(i.e., for nonresonant transport).  Furthermore, the inelastic transmission probability is negligible compared to Eq.\ \ref{eq:trans_matrix}
in that limit.

%The Green's function of the nearly isolated molecule $G_{\rm mol}$ is found by exactly diagonalizing the molecular Hamiltonian---projected onto
The molecular Green's function $G_{\rm mol}$ is found by exactly diagonalizing the molecular Hamiltonian, projected onto
a basis of relevant atomic orbitals
%---in the sequential-tunneling limit $\Sigma_{\rm T}\rightarrow 0$
\cite{Bergfield09}:
\begin{equation}
	G_{\rm mol}(E) = \sum_{\nu, \nu'} \frac{[{\cal P}(\nu) + {\cal P}(\nu')] C(\nu,\nu')}{E-E_{\nu'}+E_{\nu}+i0^+},
	\label{eq:Gmol}
\end{equation}
where $E_\nu$ is the eigenvalue associated with eigenstate $\nu$ of the molecular Hamiltonian, ${\cal P}(\nu)$ is the probability that the state $\nu$
is occupied, %which in equilibrium is given by the grand canonical ensemble, 
and $C(\nu,\nu')$ is a rank-1 matrix with elements
% in equilibrium ${\cal P}(\nu)$=$e^{-\beta(E^\nu-\mu N^\nu)}/{\cal Z}$ is the probability of the state $\nu$ being occupied with $\beta$=$1/, and 
\begin{equation}
[C(\nu,\nu')]_{n\sigma,m\sigma'} = \langle \nu | d_{n\sigma} | \nu' \rangle \langle \nu' | d^\dagger_{m\sigma'} | \nu \rangle.
	\label{eq:manybody_element}
\end{equation}
Here $d_{n\sigma}$ annihilates an electron of spin $\sigma$ on the $n$th atomic orbital of the molecule.
For linear response, ${\cal P}(\nu)$ is given by the grand canonical ensemble.  Equations  \ref{eq:Dyson2}--\ref{eq:manybody_element} imply that each molecular resonance $\nu\rightarrow\nu'$ contributes at most one transmission
channel in Eq.\ \ref{eq:trans_matrix}.

An %semi-empirical Hamiltonian %\cite{Cardamone06,Bergfield09,Castleton02,JoshUnpublished}
effective field theory of interacting $\pi$-electrons 
%that accurately describes Coulomb interactions and $\pi$-conjugation
was used to model the electronic degrees of freedom most relevant for transport.
Using symmetry principles and an electrostatic multipole expansion, an effective Hamiltonian
for the $\pi$-electrons was derived 
%\cite{JoshUnpublished}
, keeping all interaction terms up to the quadrupole-quadrupole interaction. 
The minimal set of parameters necessary to characterize the $\pi$-electron system consists of
the $\pi$-orbital on-site repulsion $U_0$, the $\pi$-orbital quadrupole moment $Q$, the nearest-neighbor $\pi$--$\pi$ hybridization $t$, and the dielectric constant $\epsilon$,
which accounts for the polarizability of the $\sigma$-electrons, which are not included explicitly in the calculation.
The values $U_0$=8.9eV, $Q=0.67e\mbox{\AA}^2$, $t=2.64\mbox{eV}$, and $\epsilon=1.63$ were determined by fitting to the gas-phase spectrum of benzene.\cite{Castleton02}
%benzene spectrum
In the molecular junction,
screening of intramolecular Coulomb interactions by the nearby metal electrodes 
\cite{Neaton06,Thygesen09} was included via the image charge method %\cite{JoshUnpublished}
, with no additional adjustable parameters.

%%%%%%%%%%%%%%%%%%%%%%%%%%%%%%%%%%%%%%
%\section{The lead-molecule coupling}
%%%%%%%%%%%%%%%%%%%%%%%%%%%%%%%%%%%%%%
\begin{figure}[tb]
	\centering
		\includegraphics[width=3.2in]{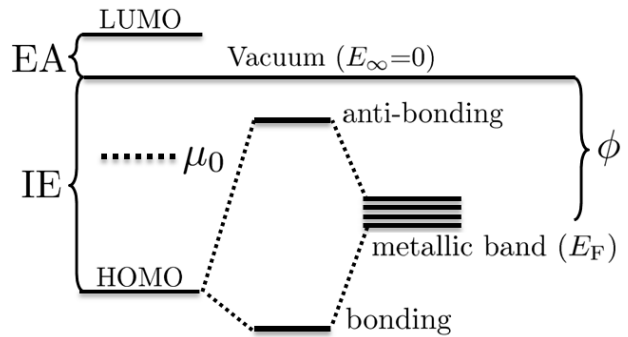}
		\caption{
Energy-level
diagram of a molecule with ionization energy IE %$E_{\rm IE}$ 
and electron affinity EA %$E_{\rm EA}$ 
bound to a metal surface/electrode with work function $\phi$.  In forming the metal-molecule bond, the \HO %and \LU 
resonance %level 
of the molecule shifts and hybridizes. 
%the \HOLU mid-gap energy $\mu_0$ is not changed in the bonding process.  
}
	\label{fig:level_structure}
\end{figure}

\section{Junction Ensemble}
In this article, we consider junctions in which two macroscopic multi-channel leads each couple to several atomic orbitals of a single molecule. 
In order to calculate the distribution of transmission eigenvalues, it is first necessary to construct a physical ensemble of junctions.
Both the lead-molecule coupling and the electrode geometry \cite{Burki03} vary over the ensemble of junctions produced in an experiment \cite{Kiguchi08}.
The lead-molecule coupling involves both {\em screening} \cite{Neaton06,Thygesen09} %of intramolecular interactions by the nearby metal electrodes
and {\em hybridization} of the molecular and metallic states, described by the matrix $\Sigma_{\rm T}$ (see Fig.\ \ref{fig:level_structure}).  
We assume both effects are dominated by the interaction of the molecule with a single
Pt atom at the tip of each electrode, as illustrated in Fig.\ \ref{fig:multimode_paper_benzene_Pt}.
Since $\Sigma_{\rm T}$ depends exponentially on the tip-molecule distance, we assume its variation is most important, and keep screening fixed over
the ensemble of junctions.  Moreover, we neglect the real part %\cite{footnote_ReSigmaT} 
of $\Sigma_{\rm T}$ and focus on the variation of the tunneling-width matrices
$\Gamma_\alpha$.  The variation of electrode geometry leads to a variation of the work function $\phi$ and density of states (DOS) 
of the leads (the latter also
contributes to the variation of $\Gamma_\alpha$.)  
%The variation of $\phi$ %the lead work function 
%has an effect similar to any variation of $\Re \Sigma_{\rm T}$.  
We thus assume the ensemble of junctions can be modeled adequately through variations of $\Gamma_\alpha$ and $\phi$
%lead work functions
only \cite{footnote_ReSigmaT}.

\subsection{Molecule-electrode Coupling}
In order to determine the
tunneling-width matrices $\Gamma_\alpha$ for a SMJ, we first consider the details of a single benzene molecule adsorbed on a Pt(111) surface.
This is the most stable Pt surface, and has been the subject of numerous investigations,
where the observed binding energy for benzene ranges between 21~kcal/mol (0.91~eV/molecule) to 29~kcal/mol
(1.26~eV/molecule) corresponding to the atop(0) and bridge(30) bonding configurations, respectively \cite{Cruz07,Morin03,Saeys02}.
As indicated schematically in Fig.\ \ref{fig:level_structure}, when a molecule binds with a metal surface, the relevant energy levels of the molecule
shift and hybridize, forming bonding and anti-bonding states. These two effects contribute to the binding energy %which may be written as 
$\Delta E_{\rm b}=\Delta E_{\rm vdW}+ \Delta E_{\rm hyb}$,
where $\Delta E_{\rm vdW}=\langle {H_{\rm mol}} \rangle-\langle \tilde{H}_{\rm mol} \rangle$ is the van der Waals energy shift 
and $\Delta E_{\rm hyb}$ is the hybridization energy.  Here $H_{\rm mol}$ is the
gas-phase molecular Hamiltonian and $\tilde{H}_{\rm mol}$ is the molecular Hamiltonian including screening from the Pt surface.
Taking the benzene-Pt distance as 2.25{\AA} and assuming the screening is dominated by the nearest Pt atom, we find
that the \HOLU gap of benzene reduces from 10.05eV in the gas-phase to 7.52eV on Pt(111) and 
$\Delta E_{\rm vdW}=0.49\mbox{eV}$.
This implies $\Delta E_{\rm hyb}\leq 0.77\mbox{eV}$.

%%%%%%%%%%%%%%%%%%%%%%%%%%%%%%%%%
%% Van der Waals and Josh's stuff
%%%%%%%%%%%%%%%%%%%%%%%%%%%%%%%%%

%%%%%%%%%%%%%%%%%%
%% Hybridization
%%%%%%%%%%%%%%%%%%
 
Since the metallic work function $\phi$ lies between the \HO and \LU resonances, hybridization occurs via the virtual exchange of an electron or
hole between the metal and the neutral molecule.
Using second-order perturbation theory, we find:
\begin{align}
\Delta E_{\rm hyb} &= \sum_{\nu\in {\cal H}_{N-1}} \int_{\mu}^\infty \frac{dE}{2\pi} \frac{\Tr{\Gamma(E)C(\nu,0_{N})}}{E-E_{0_{
N}}+E_\nu} \nonumber \\
	&+
	\sum_{\nu'\in {\cal H}_{N+1}} \int_{-\infty}^\mu \frac{dE}{2\pi} \frac{{\rm Tr} \{\Gamma(E)C(0_{N},\nu')\}}{-E-E_{0_{N}}+E_{\nu'}},
	\label{eq:delta_Ehybrid}
\end{align}
%\begin{eqnarray}
%\Delta E_{\rm hyb} &=& \sum_{\nu\in {\cal H}_{N-1}} \int_{\mu}^\infty \frac{dE}{2\pi} \frac{\Tr{\Gamma(E)C(\nu,0_{N})}}{E-E_{0_{
%N}}+E_\nu} \nonumber \\
%	%
%	&+&
%	\sum_{\nu'\in {\cal H}_{N+1}} \int_{-\infty}^\mu \frac{dE}{2\pi} \frac{{\rm Tr} \{\Gamma(E)C(0_{N},\nu')\}}{-E-E_{0_{N}}+E_{\nu'}},
%	\label{eq:delta_Ehybrid}
%\end{eqnarray}
where $\mu$ is the chemical potential of the lead metal, ${\cal H}_{N}$ is the $N$-particle molecular Hilbert space, 
and $0_{N}$ is the ground state of the $N$-particle manifold of the neutral molecule. %$N$-particle manifold.
At room temperature, the Pt DOS $g(E)$ is sharply peaked around the Fermi energy \cite{Kleber73}, allowing us to perform the energy integral in Eq.\ \ref{eq:delta_Ehybrid} using $\Gamma(E)\approx \Gamma(\varepsilon_F) Z\delta(E-\varepsilon_F)/g(\varepsilon_F)$, where we take $Z$=+4 for a Pt atom
and $g(\varepsilon_F)$=2.88/eV %, nearly 10$\times$ that of Au at the Fermi energy 
\cite{Kittel_KinderBook}.
The hybridization energy is thus determined by the tunneling-width matrix evaluated at the Pt Fermi level, $\Gamma(\varepsilon_F)$$\equiv$$\Gamma$.
%%%% Now discuss \mu_Pt(111)
Using $\phi_{\rm Pt(111)}$=5.93eV \cite{CRC} 
%and the chemical potential of benzene $\mu_0=(\varepsilon_{\rm IE}+\varepsilon_{\rm EA})/2=-4.06\mbox{eV}$ \cite{Mikaya82,Janousek79} 
and the chemical potential of benzene $\mu_0$=$({\rm IE}+{\rm EA})/2$=-4.06eV \cite{Mikaya82,Janousek79} 
(which is unaffected by screening),
we find $\Tr{\Gamma} \leq 21.6\mbox{eV}$ to fit the hybridization energy for benzene adsorbed on Pt.

%%%%%%%%%%%%%%%%%%%%%%%%%%%%%%%%%%%%%%%%%%%%%%%%
% Discuss bonding configuration and DOS ensemble
%%%%%%%%%%%%%%%%%%%%%%%%%%%%%%%%%%%%%%%%%%%%%%%%

%Equations\ (Eq.\ \ref{eq:bonding_energy_total}) and (Eq.\ \ref{eq:delta_Ehybrid}) constitute an important %formal 
%relationship between the tunneling-width matrix $\Gamma$ and the observable binding energy $\Delta E_{\rm b}$. 

\begin{figure}[tb]
	\centering	
	\includegraphics[width=3.3in]{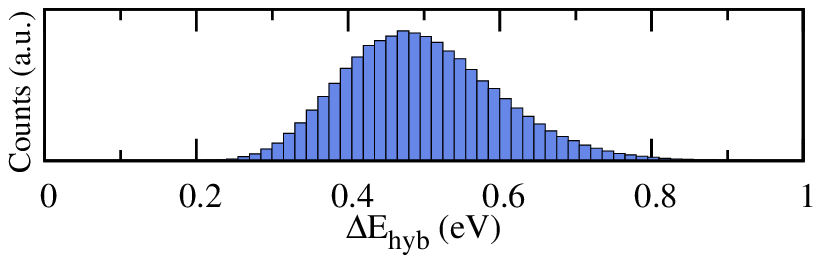}
\includegraphics[width=3.3in]{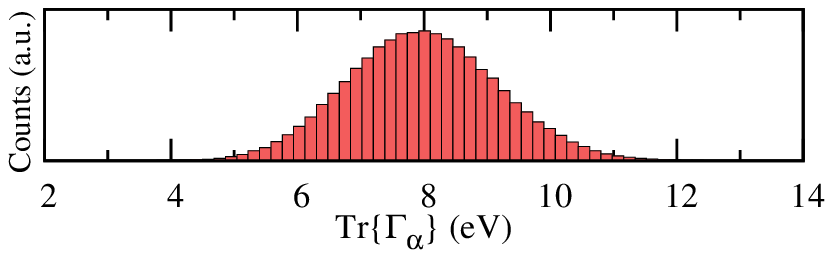}
\vspace{-.5cm}
\caption{Distributions of hybridization energy (top) and tunneling-width matrix trace (bottom) for a Pt-benzene-Pt junction.
The 99${\rm ^{th}}$-percentile values for 
$\Delta E_{\rm hyb}$ and $\Tr{\Gamma_\alpha}$
%the hybridization energy and trace 
are 0.77eV and 10.82eV, respectively.  Here $\alpha$=1,2 labels the lead-molecule contacts.}
	\label{fig:benzene_binding_and_tracegamma}
\end{figure}

\section{Results and Discussion}
\subsection{Pt-benzene-Pt Junctions}
Let us now consider lead-molecule hybridization in a Pt-benzene-Pt junction.  
Screening from two metal electrodes further reduces the \HOLU gap to 6.46eV.  Since the most favorable binding of benzene on the closest-packed Pt(111) surface gives
$\Delta E_{\rm hyb}$=$0.77$eV, we assume this is essentially an upper bound on hybridization in a SMJ, where the bonding is 
more random.
We wish to study the dependence of the transmission eigenvalue distribution on the number of bonds formed with each electrode.  
For $M$ covalent bonds between a macroscopic lead and a molecule with $P$ atomic orbitals, 
$\Gamma$ is a rank-$M$ matrix, which can be represented as %the tunneling-width matrix 
\begin{equation}
	\Gamma = \sum_{m=1}^{M} \gamma_m^\dagger \gamma_m,
	\label{eq:GammaConstruction}
\end{equation}
where $\gamma_m$ are linearly-independent real row vectors of dimension $P$,
representing linear combinations of the atomic orbitals of the molecule \cite{footnote_Tinv}.
Our approach
is to populate the elements of $\gamma_m$ from a uniform random distribution on the interval [-$A$,$A$]. 
%where $A=0.82\mbox{eV}$ is chosen according to the upper bound on $\Delta E_{\rm hybrid}$ discussed above.
The bonding ensemble corresponds to a random walk of $M$ steps in a $P$-dimensional space.
The distributions of $\Delta E_{\rm hyb}$ and $\Tr{\Gamma_\alpha}$ shown in %the top and bottom panels of 
Fig.\ \ref{fig:benzene_binding_and_tracegamma} %, respectively.
have long Gaussian tails, so the value $A=0.82\mbox{eV}$ was chosen to fix
%We therefore limit the range of elements of $\gamma_m$ such that 
the 99$^{\rm th}$-percentile of $\Delta E_{\rm hyb}$ at 0.77eV \cite{footnote_DOSfluct}.
\nocite{Stafford99}
The 99$^{\rm th}$-percentile of
$\Tr{\Gamma_\alpha}$ is 10.82eV which, per orbital, is nearly 3$\times$ the coupling found for a Au-BDT-Au junction \cite{Bergfield09}.
% 99th of Tr{Gamma}=10.8274eV
% Amax = .8155
%  

In addition to sampling a variety of bonding configurations, we assume the ensemble of junctions %produced in the experiment \cite{Kiguchi08}
samples all possible Pt surfaces.
The work function of Pt ranges from 5.93eV to 5.12eV for the (111) and (331) surfaces,
respectively \cite{CRC}, so that %and correspondingly
\begin{equation}
-{\rm 1.88eV}
\leq  \mu_{\rm Pt}-\mu_0  \leq 
	-{\rm 1.07eV},
% \tilde{E_{HOMO}}=7.283
	\label{eq:MuRange_BenzeneOnPlatinum}
\end{equation}
and we assume a uniform distribution of $\mu_{\rm Pt}$ on this interval.

%%%%%%%%%%%%%%%%%%%%%%%%%%%%%%%%%%%%%%%%%%%%%%
%\section{Benzene in platinum junction}
%%%%%%%%%%%%%%%%%%%%%%%%%%%%%%%%%%%%%%%%%%%%%%

% In code being more positive here corresponds to negative.  I.e. 1.065 would be -1.065,
% since it is below the midgap.  
%where $\mu_{\rm Pt}$ is considered with respect to $\mu_0$.

\begin{figure}[bt]
	\centering	
	\includegraphics[width=3.3in]{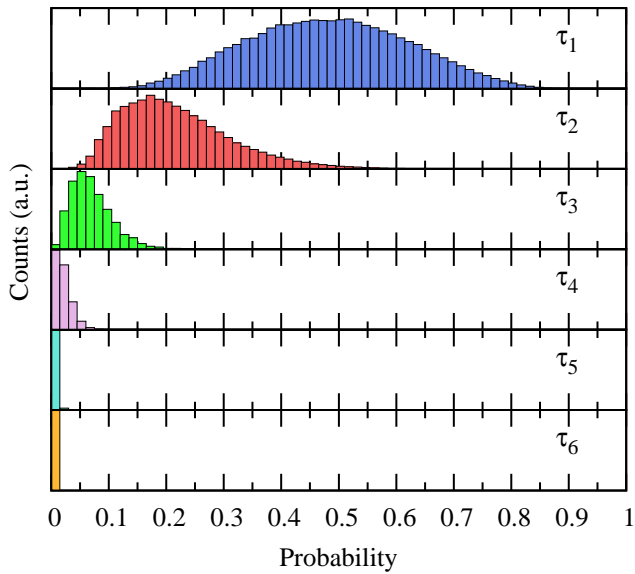}
\caption[benzene channel distribution]{Calculated transmission eigenvalue distributions for an ensemble of $24\times 10^4$ Pt--benzene--Pt junctions.
3,000 bonding configurations $\times$ 80 Pt surfaces were included.  Each lead was assumed to form
$M$=6 bonds with the molecule.
There are two dominant transmission channels arising from the two-fold degenerate \HO resonance closest to the Pt Fermi level, with a small
but experimentally resolvable third channel arising from further off-resonant tunneling.
%The total conductance of the junction $G$=$G_0 \sum_i \tau_i$, where $G_0$=$2e^2/h$.
}
	\label{fig:benzene_channel_distributions}
\end{figure}

The transmission eigenvalue distributions 
are shown in Fig.\ \ref{fig:benzene_channel_distributions} 
for an ensemble of $24\times 10^4$ Pt--benzene--Pt junctions.
Each lead was assumed to form $M=6$ bonds with the molecule.
Despite the existence of six covalent bonds between the molecule and each lead, 
there are only two dominant transmission channels, which arise from the two-fold degenerate \HO resonance closest to the Pt Fermi level. 
%(TO DO: supplementary calculation to prove this.)
There is also a small
but experimentally resolvable third channel arising from further off-resonant tunneling.
This channel is non-negligible because of the very large lead-molecule coupling $\Gamma$ in the Pt-benzene junction.  
% New sentence
Explicit calculations with smaller values of $\Tr{\Gamma}$ yielded only two non-negligible transmission eigenvalues.
Thus, 
for metals with a smaller DOS at the
Fermi level, such as Cu, Ag, or Au, junctions with benzene would be expected to exhibit only two measurable transmission channels.

\begin{figure}[bt]
	\centering	
	\includegraphics[width=3.3in]{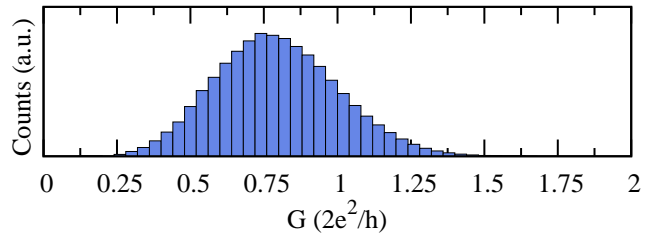}
%	\hspace{-.5cm}
\caption[benzene conductance distribution]{
Calculated conductance histogram for the same ensemble of %$24\times 10^4$ 
Pt--benzene--Pt junctions discussed in Fig.\ \ref{fig:benzene_channel_distributions}, where $G=(2e^2/h) \sum_n \tau_n$.
%10,000 bonding configurations $\times$ 80 Pt surfaces were included.  Each lead was assumed to form $M=6$ bonds with the molecule. 
%but the histogram is similar for any value of $M\geq 2$.
%The peak conductance value is somewhat less than that reported experimentally \cite{Kiguchi08}.  This discrepancy might be 
%attributable, in part, to the inclusion of a small fraction of Pt--Pt junctions in the experimental histogram.
}
	\label{fig:benzene_conductance_histogram}
\end{figure}

The calculated transmission eigenvalue distribution shown in Fig.\ \ref{fig:benzene_channel_distributions} is
consistent with the experiment \cite{Kiguchi08}, which determined the transmission eigenvalues %$\{\tau_n\}$ 
for three particular 
junctions:  $\{\tau_n\}$=$\{0.68, \, 0.40\}$, $\{0.36,\, 0.25,\, 0.10\}$, $\{0.20\}$,
where a third channel was observable only once. 
%We stress that we have not used any results of Ref.\ \onlinecite{Kiguchi08} in our simulation.  
The conductance histogram for the same ensemble of 
%$24\times 10^4$ Pt--benzene--Pt 
junctions
%10,000 bonding configurations $\times$ 80 values of $\mu_{\rm Pt}$ 
is shown in Fig.\ \ref{fig:benzene_conductance_histogram}.
%Each lead was assumed to form
%$M=6$ bonds with the molecule, but the histogram is similar for any value of $M\geq 2$.
The peak conductance value is some 20\% less than that reported experimentally \cite{Kiguchi08}.  This discrepancy might be
attributable, in part, to the inclusion of a small fraction of Pt--Pt junctions in the experimental histogram.

In addition to the ensemble of junctions shown in Fig.\ \ref{fig:benzene_channel_distributions}, we also investigated ensembles of 
junctions with $M_\alpha=1,\ldots,6$, including the case $M_1\neq M_2$.  Consistent with the discussion in Refs.\ \cite{Solomon06b,Kiguchi08},
we find that the total number of nonzero transmission eigenvalues is $M_{\rm min}=\min\{M_1,M_2\}$.
However, whenever $M_{\rm min}\geq 2$ there are always two dominant transmission channels, and the total transmission probability does not
increase appreciably beyond $M_{\rm min}=2$.

%%%%%%%%%%%%%%%%%%%%%%%%%%%%%%%%%%%%%%%%%%%%%%%%%%%%%%%%%%%%%%%%%%%%%%%%%%%%%%%%
% New paragraph on overlap of dominant transmission channels with HOMO resonance (revised 9-13-2010)
%%%%%%%%%%%%%%%%%%%%%%%%%%%%%%%%%%%%%%%%%%%%%%%%%%%%%%%%%%%%%%%%%%%%%%%%%%%%%%%%

The above analysis demonstrates that the two dominant transmission channels evolve from the two-fold degenerate
\HO %HOMO 
resonance in Eq.\ \ref{eq:Gmol} as the lead-molecule coupling $\Sigma_{\rm T}$ is turned on. 
For finite $\Sigma_{\rm T}$, the %various 
poles of $G_{\rm mol}$ are mixed by Dyson's equation (Eq.\ \ref{eq:Dyson2}), making it problematic 
to decompose the transmission {\em eigenvalues} into separate contributions from each molecular resonance \cite{Solomon06b}.
Alternatively, the projections of the transmission {\em eigenvectors} onto the molecular resonances can be computed \cite{Heurich02}. 
%the absolute squares of which can be interpreted as their contributions %of each molecular resonance 
%to each transmission channel.
Because an ``extended molecule'' is often used %\cite{Heurich02} 
in density-functional calculations to account for charge transfer between the
molecule and electrodes, it is difficult if not impossible 
to interpret these contributions in terms of the resonances of the molecule itself \cite{Heurich02}.
Since charging effects in SMJs are well-described in our many-body theory \cite{Bergfield09}, there is no need to utilize an ``extended molecule,''
so the projections of the transmission eigenvectors onto the molecular resonances can be determined unambiguously (see Methods).
We find that the mean-square projections of the first and second transmission channels in Fig.\ \ref{fig:benzene_channel_distributions}
onto the benzene \HO resonance
are 87\% and 71\%, respectively,
confirming the conclusion that these eigenchannels correspond to tunneling primarily
through the \HO resonance.

%%%%%%%%%%%%%%%%%%%%%%%%%%%%%%%%%%%%%%%
% Nondegenerate MO = 1 dominant channel
%%%%%%%%%%%%%%%%%%%%%%%%%%%%%%%%%%%%%%%

%%%%%%%%%%%%%%%%%%%%%%%%%%%%%%%%%%%%%%%%%%%%%%%%%
%%%%%%%%%%%%%%%%%%%%%%%%%%%%%%%%%%%%%%%%%%%%%%%%%
%%%%%%%%%%%%%%%%%%%%%%%%%%%%%%%%%%%%%%%%%%%%%%%%%
%\section{1,3-Butadiene on Pt}
%U0 -> E_4 = -10.158eV, E_3=-5.538eV; Gap=2*4.62eV
%Single sphere -> E_4 = -10.680eV, E_3 = -7.4879eV; Gap=2*3.1921eV
%Double sphere-> E_4=-11.153eV, E_3=-8.5969eV; Gap=2*2.5561eV 
%The 99th percentile of Trace Gamma = 5.468eV
% -1.699eV < \mu_Pt - \mu_0 < -.889eV

\begin{figure}[bt]
	\centering	
\includegraphics[width=3.3in]{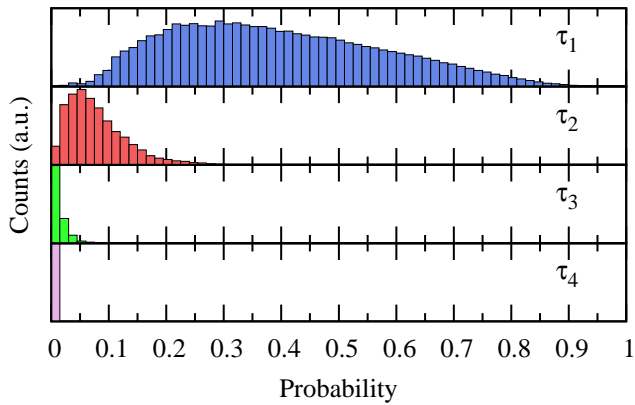}
\caption[Butadiene plot]{Transmission eigenvalue distributions for an ensemble of 
$24\times 10^4$ Pt--butadiene--Pt junctions.
3,000 bonding configurations $\times$ 80 Pt surfaces were included.  Each lead was assumed to form
$M=4$ bonds with the molecule.
The linear butadiene (${\rm C_4H_6}$) molecule lacks the orbital degeneracy of the benzene ring and consequently exhibits
only 1 dominant channel. A second channel due to further off-resonant tunneling may be experimentally resolvable.  The total conductance distribution peaks around $\sim$0.4G$_0$.
}
	\label{fig:butadiene_histogram}
\end{figure}

\subsection{Pt-butadiene-Pt Junctions}
To test our hypothesis that the number of dominant transmission channels is limited by the degeneracy of the most relevant molecular orbital, 
we have investigated transmission eigenvalue distributions for Pt--butadiene--Pt junctions.  Butadiene is a linear
conjugated polymer, lacking the discrete (six-fold) rotational symmetry of benzene. Since the molecular orbitals of butadiene are nondegenerate, 
we anticipate a single dominant transmission channel.
Using the same per-orbital hybridization as for benzene gives  $\Tr{\Gamma_\alpha}\leq 7.21 \mbox{eV}$.  The image charge method gives 
$\Delta E_{\rm vdW}=0.52\mbox{eV}$ and we find $\Delta E_{\rm hyb}\leq 0.59\mbox{eV}$, so that
$\Delta E_{\rm b} \leq 1.12\mbox{eV}\equiv 107.7\mbox{kJ/mol}$, in line with
existing values reported in the literature \cite{Valcarcel04}.  
The range of Pt work functions for all possible Pt surfaces gives a chemical potential range of
\begin{equation}
 -1.70\mbox{eV} \leq \mu_{\rm Pt} - \mu_0 \leq -0.89\mbox{eV}.
\label{eq:MuRange_ButadieneOnPlatinum}
\end{equation}
Despite forming four bonds with each electrode, it is evident from Fig.\ \ref{fig:butadiene_histogram} that the Pt--butadiene--Pt junction has a single
dominant transmission channel. 
The mean-square projection of this transmission channel onto the non-degenerate butadiene \HO resonance is 80\% (see Methods).

\section{Conclusions}
%In conclusion, we find that the number of dominant transmission channels
%in a SMJ is equal to the degeneracy of the molecular orbital closest to the metal Fermi level.
%Since molecules can possess only discrete spatial symmetries, 
%molecular orbitals can have at most two-fold orbital degeneracy (barring accidental degeneracies).
%SMJs can thus have only one or two dominant transmission channels, depending on whether the molecular 
%orbital involved in transport is nondegenerate (e.g., butadiene) or doubly degenerate (e.g., benzene).  Despite their larger molecular Hilbert space, the number of transmission channels in SMJs is more limited than in single-atom contacts because molecules are less symmetrical than atoms. 

In conclusion, we find that the number of dominant transmission channels in a SMJ is equal to the
degeneracy of the molecular orbital closest to the metal Fermi level. 
Transmission eigenvalue distributions were calculated for Pt-benzene-Pt and Pt-butadiene-Pt
junctions using realistic state-of-the-art many-body techniques.
In both cases, transmission occurs primarily through the \HO resonance, which lies closest
to the Pt Fermi level, resulting in two dominant transmission channels for benzene 
(2-fold degenerate {\sc HOMO}) and a single dominant transmission channel for butadiene (non-degenerate
{\sc HOMO}).  Our results for the transmission channel distributions of 
Pt-benzene-Pt junctions (see Fig.\ \ref{fig:benzene_channel_distributions}) are in quantitative agreement with experiment \cite{Kiguchi08}.

Despite the larger number of states available for tunneling transport in SMJs,
we predict that the number of transmission channels is typically more limited than in single-atom contacts
because molecules are less symmetrical than atoms.  
Nonetheless, certain highly-symmetric molecules exist that should permit several dominant transmission channels.
For example, the C$_{60}$ molecule possesses icosahedral symmetry,
and has a 5-fold degenerate \HO resonance and 3-fold degenerate \LU resonance \cite{Iwasa03}.
C$_{60}$-based SMJs with gold electrodes have been fabricated, and 
shown to exhibit fascinating electromechanical \cite{Park00} and 
spin-dependent \cite{Natelson04} transport properties.
For Pt-C$_{60}$-Pt junctions, we predict {\em five dominant transmission channels} stemming from
the C$_{60}$ \HO resonance, which lies closest to the Pt Fermi level \cite{Lichtenberger1991,Kroto1991}.

\section{Methods}

%\section{Supplementary information}
\subsection{Decomposing transmission channels in terms of molecular resonances}

%\label{sec:sup_info}

In order to understand how transport in a SMJ is determined by the chemical properties of the molecule, it would be desirable to 
express the transmission eigenchannels in terms of molecular resonances \cite{Heurich02,Solomon06b}.
The transmission eigenvalues $\tau_n$ and eigenvectors $|n\rangle$ are solutions of the equation
\begin{equation}
T |n\rangle = \tau_n |n\rangle,
\label{eq:T_eigenproblem}
\end{equation}
where $T$ is the transmission matrix given by Eq.\ \ref{eq:trans_matrix}.  By definition, $|n\rangle$ is a linear combination of the atomic orbitals of the molecule.  In an effective single-particle model \cite{Heurich02}, 
$|n\rangle$ may also be expressed as a linear combination of {\em molecular orbitals} $|\phi_j\rangle$
\begin{equation}
|n\rangle = \sum_j \alpha_{n}^{\;j} |\phi_j\rangle.
\label{eq:heurich}
\end{equation}
Here $|\alpha_{n}^{\;j}|^2$
%The absolute square of the projection $\alpha_{n}^j=\langle n|\phi_j\rangle$ 
can be identified
as the contribution of the $j$th molecular orbital to the $n$th transmission channel \cite{Heurich02},
which
%$|\alpha_{n}^{\;j}|^2$ 
can be conveniently expressed in terms of the projection operator $\hat{P}_j=|\phi_j\rangle\langle\phi_j|$ as
$|\alpha_{n}^{\;j}|^2=\langle n|\hat{P}_j|n\rangle$.

In the many-body problem, %\cite{Bergfield09}
there is no orthonormal set of ``molecular orbitals''; rather
each molecular resonance of energy $E_{\nu'}-E_\nu$ corresponds to a transition $\nu \rightarrow \nu'$ between an $N$-body and an $(N+1)$-body
molecular eigenstate [see Eqs.\ \ref{eq:Gmol} and \ref{eq:manybody_element}].
%Nonetheless, it is possible to define a projection operator onto each molecular resonance
The projection operator onto a molecular resonance is
\begin{equation}
\hat{P}_{\nu\rightarrow\nu'}\equiv \frac{C(\nu,\nu')}{\Tr{C(\nu,\nu')}},
\label{eq:projection_operator}
\end{equation}
where $C(\nu,\nu')$ is given by Eq.\ \ref{eq:manybody_element}.
%where $C(\nu,\nu')$ is given by Eq.\ (4) in the article.
The absolute square projection of the $n$th transmission eigenvector onto the resonance $\nu \rightarrow \nu'$ is given by
\begin{equation}
%\left|\alpha_n^{\;\nu\rightarrow\nu'}\right|^2= \langle  n | \hat{P}_{\nu\rightarrow\nu'} |n\rangle.
|\alpha_n^{\;\nu\rightarrow\nu'}|^2= \langle  n | \hat{P}_{\nu\rightarrow\nu'} |n\rangle.
\label{eq:projection_amplitude_squared}
\end{equation}
A {\em necessary condition} to identify an eigenchannel $|n \rangle$
with transmission through
a particular molecular resonance %$\nu \rightarrow \nu'$ 
is $|\alpha_n^{\;\nu\rightarrow\nu'}|^2\approx 1$.

The above procedure is in principle straightforward to implement in an effective single-particle model based  on density 
functional theory (DFT)\cite{Heurich02}.
However, in practice,
%the absolute squares of which can be interpreted as their contributions %of each molecular resonance 
%to each transmission channel.
an ``extended molecule'' must be used %\cite{Heurich02} 
in DFT calculations to account for charge transfer between the molecule and electrodes. 
This is because current implementations of DFT fail
to account for the {\em particle aspect} of the electron  \cite{Toher05,Burke06,Datta06,Geskin09}, i.e., the strong
tendency for the electric charge on the molecule within the junction to be quantized in integer multiples of the electron charge $e$.
Analyzing transport in terms of extended molecular orbitals has unfortunately proven problematic.
%in terms of the resonances of the molecule itself. % \cite{Heurich02}.
For example, the resonances of the extended molecule in Ref.\ \citenum{Heurich02} apparently accounted for less than 9\% of the current through the junction.

Since charging effects in SMJs are well-described in our many-body theory \cite{Bergfield09}, there is no need to utilize an ``extended molecule,''
so the projections of the transmission eigenvectors onto the molecular resonances can be determined directly from Eq.\ \ref{eq:projection_amplitude_squared}.

\subsubsection{Benzene resonances}
%\label{sec:benzene_res}

The neutral ground state of benzene is nondegenerate, while the \HO and \LU resonances %of benzene 
are both doubly degenerate due to the (six-fold) 
rotational symmetry of the molecule.  To be consistent with the discussion of Ref.\ \citenum{Kiguchi08}, the additional two-fold spin degeneracy of
each resonance is considered implicit.  

We define the following projection operators:
\begin{align}
\hat{P}_{\rm HOMO} & \equiv  \sum_{\nu \in 0_5} \hat{P}_{\nu\rightarrow 0_6},  \label{eq:P_homo} \\	
\hat{P}_{\rm LUMO} & \equiv  \sum_{\nu' \in 0_7} \hat{P}_{0_6\rightarrow \nu'}, \label{eq:P_lumo} \\
\hat{P}_\perp & \equiv  \mathbf{1} - \hat{P}_{\rm HOMO} -\hat{P}_{\rm LUMO},\label{eq:P_perp} 
\end{align}
where $0_N$ is the ground state with $N$ $\pi$-electrons and $\mathbf{1}$ is the six-dimensional unit matrix in the space of $\pi$-orbitals.
$\hat{P}_{\rm HOMO}$ and $\hat{P}_{\rm LUMO}$ are projection operators onto the two-dimensional
subspaces spanned by the \HO and \LU resonances, 
respectively.  Eqs.\ \ref{eq:P_homo}--\ref{eq:P_perp} define a complete, orthogonal set of projection operators in six-dimensions with the properties
\begin{align}
\sum_j \hat{P}_j &=  \mathbf{1},
\label{eq:completeness} \\
\hat{P}_i \hat{P}_j &=  \hat{P}_i \delta_{ij}.
\label{eq:orthogonality}
\end{align}
In particular, Eq.\ \ref{eq:orthogonality} implies that the \HO and \LU subspaces of benzene are orthogonal.  The absolute square of the projection of the n$^{\rm th}$ transmission eigenvector onto the subspace spanned by $\hat{P}_j$ is 
\begin{equation}
	|\alpha_n^{\; j}|^2 = \langle n|\hat{P}_j|n\rangle.
\end{equation}
These coefficients satisfy the condition 
\begin{equation}
\sum_j |\alpha_n^{\; j}|^2 = 1,
\end{equation}
where the sum runs over $j$={\sc HOMO, LUMO}, $\perp$.

%In analogy with Eq.\ (Eq.\ \ref{eq:heurich}), we can write 
%\begin{equation}
%|n\rangle = \sum_j \alpha_n^{\; j} \hat{P}_j |n\rangle,
%\label{eq:linear_comb}
%\end{equation}
%where 

\begin{figure}[ptb] %[bth]
	\centering	
\includegraphics{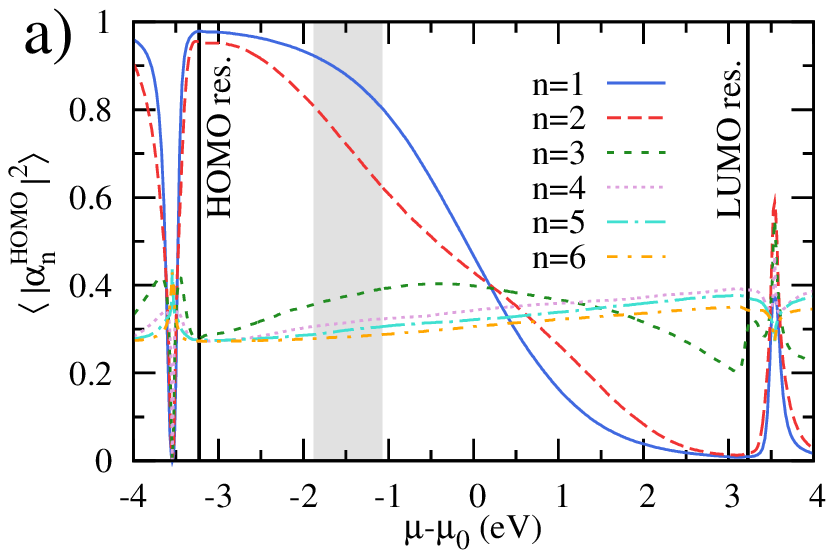}
\includegraphics{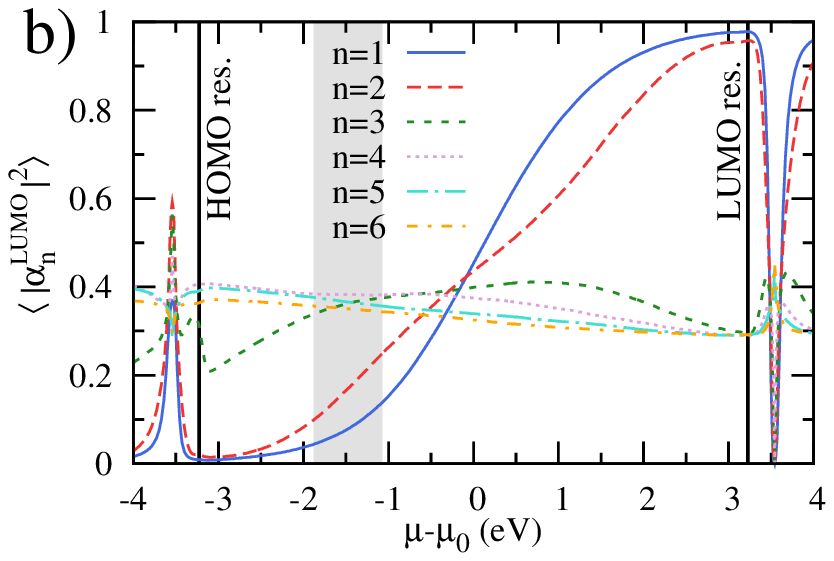}
\includegraphics{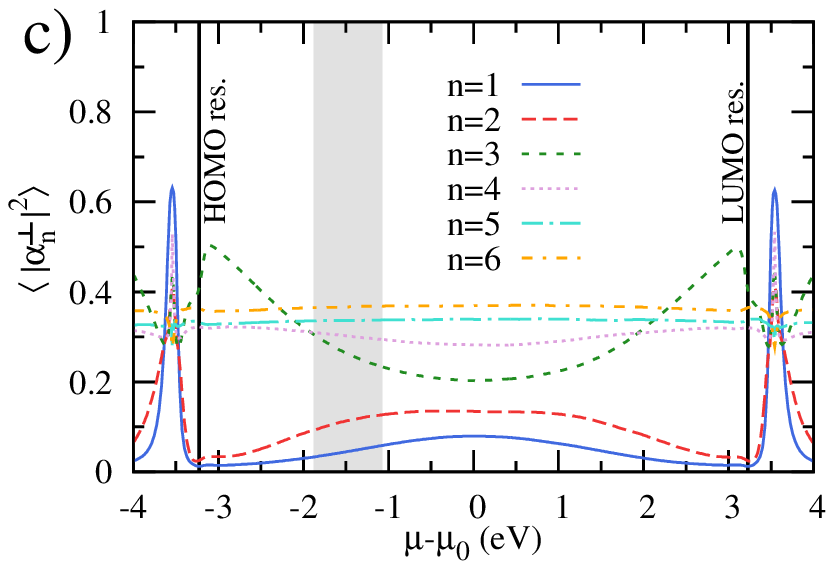}
\caption[Benzene channel overlaps]{
The mean-square projections $|\alpha_n^{\; j}|^2=\langle n|\hat{P}_j|n\rangle$ of the transmission eigenvectors
onto the two-dimensional subspaces (a) spanned by the benzene \HO %{\sc HOMO} 
resonance; (b) spanned by the benzene \LU resonance; and (c) orthogonal to the \HO and \LU subspaces,
for the same ensemble of %$24\times 10^4$ 
Pt--benzene--Pt junctions discussed in
Fig.\ \ref{fig:benzene_channel_distributions}. %Note that 
For the range of possible chemical potentials of Pt electrodes, -1.88eV$
\leq  \mu_{\rm Pt}-\mu_0  \leq$ -1.07eV indicated by the grey boxes in each subfigure, the first two channels have very strong overlap with the \HO resonance: 87\% and 71\%, respectively.  
}
	\label{fig:benzene_channel_overlaps}
\end{figure}

Fig.\ \ref{fig:benzene_channel_overlaps} shows the mean-square projections $\langle|\alpha_n^{\; j}|^2\rangle$ %=\langle n|\hat{P}_{\sc HOMO}|n\rangle$ 
of the transmission eigenvectors onto (a) %the two-dimensional subspace spanned by 
the benzene \HO resonance; (b) the benzene \LU resonance; and (c) the two-dimensional subspace orthogonal to both the \HO and \LU 
resonances, as a function of electrode chemical potential
for the same ensemble of %$24\times 10^4$ 
Pt--benzene--Pt junctions discussed in Fig.\ \ref{fig:benzene_channel_distributions}.
We find that the mean-square projections of the first and second transmission channels onto the benzene \HO resonance
are 87\% and 71\%, respectively,
confirming the conclusion that these eigenchannels correspond to tunneling primarily
through the \HO resonance.

Midway between the \HO and \LU resonances at $\mu=\mu_0$, 
the first two transmission channels have mean-square projections of $\approx 0.5$ onto both the \HO and \LU resonances,
consistent with the expectation that the \HO and \LU resonances should contribute equally to transmission.
The remaining channels do not have negligible overlap with the \HO resonance, but instead cluster around $\langle |\alpha_n^{\rm HOMO}|^2\rangle \sim 0.3$.
The transmission channels with $\langle \tau_n\rangle \ll 1$ correspond to contributions from several far off-resonant poles, each of which has
some overlap with the \HO resonance due to the overcompleteness of the projectors $\hat{P}_{\nu\rightarrow\nu'}$ \cite{footnote_overcompleteness}.  
%\footnote{Note that there are 600 poles
%of Eq.\ (Eq.\ \ref{eq:Gmol}) for a neutral benzene molecule consistent with particle-number and $S_z$ selection rules, although some of these transitions
%are forbidden by the $S^2$ selection rule, vastly more than the six atomic orbitals in the basis set.} 
They are essentially random unit
vectors in the six-dimensional space of benzene $\pi$-orbitals, whose mean-square overlap with the two-dimensional \HO subspace should be
$\langle |\alpha_n^{\rm HOMO}|^2\rangle =2/6=1/3$.

The first two transmission channels only have an appreciable overlap with the subspace orthogonal to the \HO and \LU resonances in the vicinity of
a pronounced dip in the transmission spectrum at $\mu-\mu_0 \approx \pm 3.5\mbox{eV}$. % caused by destructive quantum interference.

\pagebreak

\subsubsection{Butadiene resonances}
\label{sec:butadiene_res}

\begin{figure}[htb] %[bt]
	\centering	
\includegraphics[width=3in]{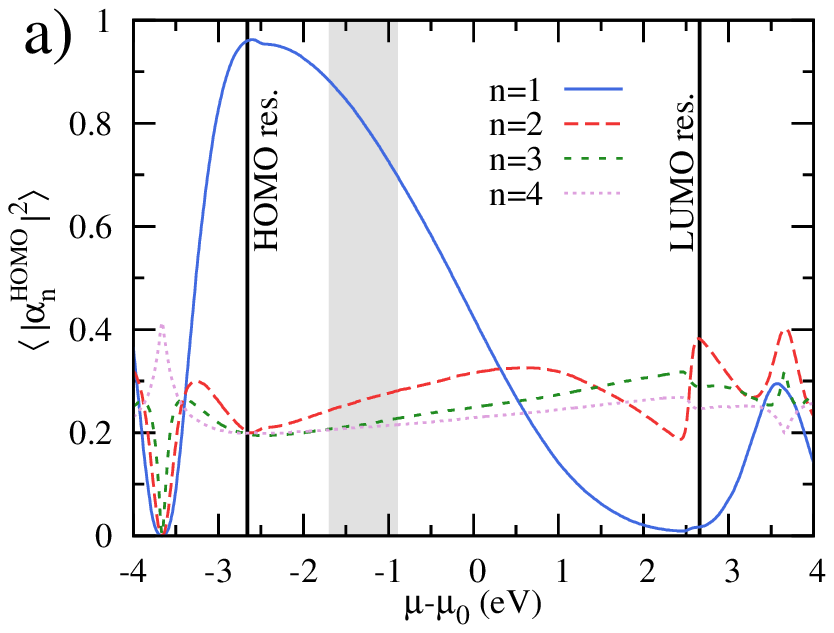}
\includegraphics[width=3in]{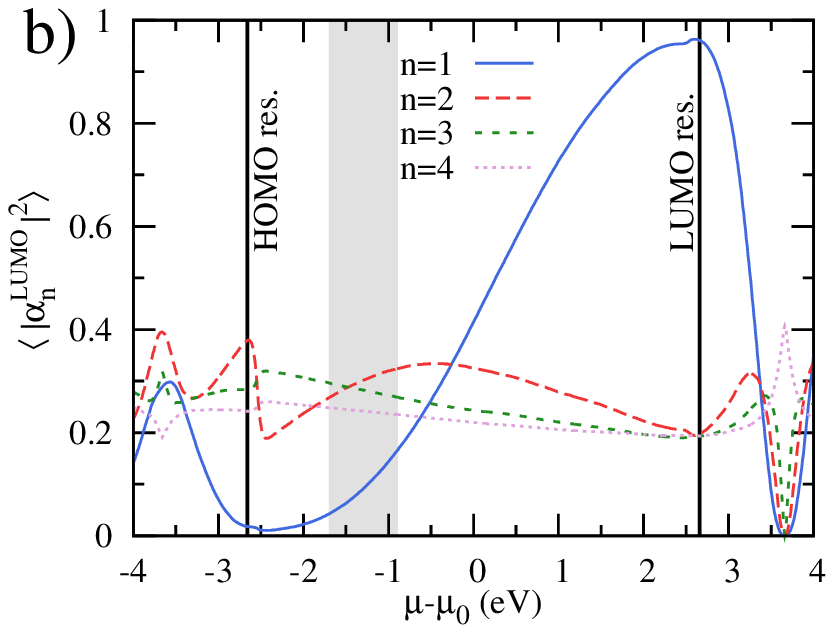}
\caption[Butadiene channel overlaps]{
The mean-square projections $|\alpha_n^{\; j}|^2=\langle n|\hat{P}_j|n\rangle$ of the transmission eigenvectors
onto (a) the %nondegenerate 
butadiene \HO resonance; and (b) the %nondegenerate 
butadiene \LU resonance,
for the same ensemble of %$24\times 10^4$ 
Pt--butadiene--Pt junctions discussed in
Fig.\ \ref{fig:butadiene_histogram}.
The dominant channel has a strong overlap (80\% mean-square) with the \HO resonance for the range of possible chemical potentials of Pt electrodes, -1.70eV$ \leq \mu_{\rm Pt} - \mu_0 \leq$ -0.89eV, indicated on each subfigure by a solid grey box.
}
	\label{fig:butadiene_channel_overlaps}
\end{figure}

The neutral ground state of butadiene is nondegenerate, and the \HO and \LU resonances have no orbital degeneracy.
The projection operators onto the \HO and \LU resonances of butadiene are
\begin{align}
\hat{P}_{\rm HOMO} & \equiv  \hat{P}_{0_3\rightarrow 0_4}, \label{eq:P_homo2} \\
\hat{P}_{\rm LUMO} & \equiv  \hat{P}_{0_4\rightarrow 0_5}, \label{eq:P_lumo2} 
\end{align}
respectively.  
Fig.\ \ref{fig:butadiene_channel_overlaps} shows 
the mean-square projections %$\langle|\alpha_n|^2\rangle$ %=\langle n|\hat{P}_{\sc HOMO}|n\rangle$ 
of the transmission eigenvectors onto the nondegenerate butadiene
\HO and \LU resonances as a function of electrode chemical potential
for the same ensemble of %$24\times 10^4$ 
Pt--butadiene--Pt junctions discussed in
Fig.\ \ref{fig:butadiene_histogram}.
In this case, the single dominant channel has a strong overlap 
with the nondegenerate \HO resonance:
$\langle |\alpha_1^{\rm HOMO}|^2\rangle =0.80$ when averaged over the range -1.70eV$\leq \mu_{\rm Pt}-\mu_0 \leq$~-0.89eV, %given by Eq.\ (Eq.\ \ref{eq:MuRange_ButadieneOnPlatinum}), 
while the far off-resonant channels with 
$\langle \tau_n\rangle \ll 1$ have $\langle |\alpha_n^{\rm HOMO}|^2\rangle \sim 1/4$, as expected based on the arguments given above.

%\begin{suppinfo}
%We present an analysis of the decomposition of the transmission channels in terms of molecular resonances.
%\end{suppinfo}

%%%%%%%%%%%%%%%%%%%%%%%%%%%%%%%%%%%%%%%%%%%%
% Bibliography
%%%%%%%%%%%%%%%%%%%%%%%%%%%%%%%%%%%%%%%%%%%%
%\bibliography{refs}

\begin{mcitethebibliography}{40}
\bibitem[foo()]{footnote_Tmatrix}
Often, the transmission matrix is expressed in the basis of propagating states
  in the leads $\tilde{T}=t t^\dagger$, where $t=i\gamma_1 G \gamma_2^\dagger$
  and $\Gamma_\alpha = \gamma_\alpha^\dagger \gamma_\alpha$, rather than in the
  molecular basis. It is straightforward to show that $T$ and $\tilde{T}$ have
  the same nonzero eigenvalues: Suppose $\tilde{T}|\tilde{n}\rangle =
  \tau_n|\tilde{n}\rangle$. Let $|n\rangle = \gamma_1^\dagger
  |\tilde{n}\rangle$. Then $T|n\rangle = \gamma_1^\dagger
  \tilde{T}|\tilde{n}\rangle = \tau_n |n\rangle$.\relax
\mciteBstWouldAddEndPunctfalse
\mciteSetBstMidEndSepPunct{\mcitedefaultmidpunct}
{}{\mcitedefaultseppunct}\relax
\EndOfBibitem
\bibitem[foo()]{footnote_ReSigmaT}
The variation of $\phi$ has an effect similar to any variation of $\Re
  \Sigma_{\rm T}$.\relax
\mciteBstWouldAddEndPunctfalse
\mciteSetBstMidEndSepPunct{\mcitedefaultmidpunct}
{}{\mcitedefaultseppunct}\relax
\EndOfBibitem
\bibitem[foo()]{footnote_Tinv}
$\Gamma$ is a real symmetric matrix for systems with time-reversal
  invariance.\relax
\mciteBstWouldAddEndPunctfalse
\mciteSetBstMidEndSepPunct{\mcitedefaultmidpunct}
{}{\mcitedefaultseppunct}\relax
\EndOfBibitem
\bibitem[foo()]{footnote_DOSfluct}
Very similar results for transport are obtained if the 95${\rm ^{th}}$
  percentile is used. Hybridization can be enhanced in the SMJ due to
  mesoscopic fluctuations \cite{Stafford99} in the lead DOS.\relax
\mciteBstWouldAddEndPunctfalse
\mciteSetBstMidEndSepPunct{\mcitedefaultmidpunct}
{}{\mcitedefaultseppunct}\relax
\EndOfBibitem
\bibitem[foo()]{footnote_overcompleteness}
Note that there are 600 poles of the molecular Green's function (3) for a
  neutral benzene molecule, consistent with particle-number and $S_z$ selection
  rules (although some of these transitions are forbidden by the $S^2$
  selection rule), vastly more than the six atomic orbitals in the basis
  set.\relax
\mciteBstWouldAddEndPunctfalse
\mciteSetBstMidEndSepPunct{\mcitedefaultmidpunct}
{}{\mcitedefaultseppunct}\relax
\EndOfBibitem
\end{mcitethebibliography}

\providecommand*\mcitethebibliography{\thebibliography}
\csname @ifundefined\endcsname{endmcitethebibliography}
  {\let\endmcitethebibliography\endthebibliography}{}

\end{document}